\documentstyle[aps,pre,epsf,amsfonts]{revtex}
\twocolumn

\begin{document}

\title{Brownian motion exhibiting absolute negative mobility}
\author{Ralf Eichhorn, Peter Reimann, and Peter H\"anggi}

\address{Universit\"at Augsburg, Theoretische Physik I,
Universit\"atsstr. 1, D-86135 Augsburg, Germany}

\maketitle

\begin{abstract}
We consider a single Brownian particle in a spatially 
symmetric, periodic system far from thermal equilibrium.
This setup can be readily realized experimentally.
Upon application of an external static force $F$, 
the average particle velocity is negative for $F>0$ and 
positive for $F<0$ (absolute negative mobility).
\end{abstract}
\vspace{2mm}

When a system at rest is perturbed by a static force, we expect
that it responds by moving into the direction of that force.
The rather surprising opposite behavior in the form of a permanent
motion against a (not too large) 
static force of whatever direction is called
{\em absolute negative mobility} (ANM) \cite{ban72,rei99}.
If the unperturbed system is at thermal equilibrium then ANM is 
impossible since it could be exploited to construct a perpetuum 
mobile of the second kind.
Legendary but quite complex non-equilibrium systems which do exhibit 
ANM are ``donkeys'' \cite{cle01}.
In this Letter we demonstrate that even a simple, structureless
Brownian particle can exhibit ANM away from thermal equilibrium
in a setup which can be easily realized experimentally
--- a scenario hitherto commonly considered as impossible.

Going {\em in medias res}, let us
consider the overdamped 2-dim. Brownian motion
\begin{eqnarray}
\eta\,\dot x(t) & = & -\partial_x V(x(t),y(t))+\xi_x(t)\nonumber\\
\eta\,\dot y(t) & = & -\partial_y V(x(t),y(t))+\xi_y(t)+F\ ,
\label{10}
\end{eqnarray}
where $\eta$ is the viscous friction coefficient
and $V(x,y)$ is the hard-wall potential from Fig. 1a,
confining the Brownian motion to a ``corridor'' along the $y$-axis
with {\em symmetric}, {\em periodic} obstacles.
Further, $\xi_x(t)$ are thermal fluctuations, modeled by unbiased Gaussian
white noise with 
$\langle\xi_x(t)\,\xi_x(s)\rangle = 2\eta\, k_B T\,\delta(t-s)$,
where
$k_B$ denotes Boltzmann's constant and $T$ the temperature.
If $\xi_y(t)$ is an independent, second thermal white noise 
source then (\ref{10}) is an 
equilibrium system and the average particle current
\begin{equation}
\langle\dot y\rangle := 
\left\langle \lim_{t\to\infty}\frac{y(t)-y(t_0)}{t-t_0}\right\rangle
\label{20}
\end{equation}
always runs into the direction of the static force $F$.
Fig. 1b depicts the corresponding 
$\langle\dot y\rangle$-$F$-characteristics
for the simplest non equilibrium model, namely a 
{\em symmetric} dichotomous
noise $\xi_y(t)$, with the promised ANM as its most outstanding feature.

Our first remark is that 
the so-called ratchet effect \cite{rei00} is characterized
by a current $\langle\dot y\rangle$ which is non-zero for $F=0$ and does
not change its direction within an entire neighborhood of $F=0$.
This effect thus inevitably involves some kind of 
asymmetry (for $F=0$), whereas our present system is
perfectly symmetric.
Second, so-called {\em differential} negative mobility (or resistance)
\cite{bal95} is typified by a negative slope of the
$\langle\dot y\rangle$-$F$-characteristics
{\em away} from $F=0$.
Thus, in both cases the salient feature of ANM is absent, namely
a current $\langle\dot y\rangle$ which is always opposite to the 
(not too large) force $F$, independently of whether $F$ is positive 
or negative.
Finally, we mention that ANM has also been observed in 
semiconductor devices \cite{ban72}
and in models for coupled Brownian motors \cite{rei99,cle01}.
However, in the first case it has an entirely quantum
mechanical origin and in the second case it is a genuine 
collective effect, without leaving room for any kind of
{\em classical, single-particle} counterpart.
Accordingly, the respective physical mechanisms are 
completely different from ours.

Returning to the ANM in Fig. 1b, its origin can be understood as follows:
Consider a (moderately large) time interval $\tau$ during which the 
dichotomous noise $\xi_y(t)$ in (\ref{10}) is constant
and $F_{tot}:=\xi_y(t)+F>0$.
When starting in one of the ``corners''
between the right ``corridor wall''
and any of the adjacent obstacles (see Fig. 1a),
the particle first closely follows the right ``corridor wall''
until it hits the next obstacle, 
then ``slides down the back'' of that obstacle, 
and afterwards performs a ``free fall'' in the $y$-direction.
Since the lateral extension of the obstacles $b$ exceeds half the 
corridor width $B/2$ the particle then hits
with a high probability $q$ the next obstacle on its way and ends up
by being trapped in the corresponding (left) ``corner''. 
In order to avoid this trap, the particle has to thermally
diffuse at least over a distance $b-(B-b)=2b-B$ 
in the positive $x$-direction
during its ``free fall'' in the $y$-direction.
With {\em increasing} force $F_{tot}$, the available time and therefore
the probability $p:=1-q$
of such a diffusive displacement {\rm decreases},
implying that the particle travels on the average
a {\em shorter} distance along the $y$-axis during the time $\tau$.
Since an analogous consideration applies for time-intervals
$\tau$ with $F_{tot}=\xi_y(t)+F<0$,
the particle motion 
on the average acquires a bias in the direction opposite to the 
static force $F$, i.e. it exhibits ANM.

In order to quantify this argument, we first note that the 
above mentioned probability $p$ of avoiding a trap (for $F_{tot}>0$)
can be approximated as \cite{f1}
\begin{equation}
p(F_{tot})=\frac{1}{2}-\frac{1}{2}\,\mbox{erf}
\left(\frac{2b-B}{\sqrt{2Lk_BT}}\sqrt{F_{tot}}\right)
\label{30}
\end{equation}
where $\mbox{erf}(x):=2\pi^{-1/2}\int_0^x e^{-y^2}\, dy$.
With probability $p$, a particle thus covers in addition to 
the ``basic distance'' of approximately $3L/2$
another period $L$ (see Fig. 1a).
It then avoids the second
trap on its way with approximately the same (relative)
probability $p$ as in (\ref{30}), i.e. a second period $L$
is covered with (absolute) probability $p^2$ etc., see Fig. 1a.
If the maximal traveling distance (avoiding all traps) 
is of the
form $(3/2+N)L$ with $N\in{\Bbb N}$, the average traveling distance
$\Delta y(\tau,F_{tot})$ thus follows as $L[3/2+p+p^2+...+p^N]$.
Neglecting that the ``free traveling speed'' $v_y:=F_{tot}/\eta$ 
is slightly reduced when the particle ``slides down the back'' of an obstacle, 
we obtain $(3/2+N)L=v_y\tau$ and hence
\begin{equation}
\Delta y (\tau,F_{tot}) = L\left\{\frac{1}{2}+
\frac{1-[p(F_{tot})]^{\frac{F_{tot}\tau}{\eta\, L}-\frac{1}{2}}}{1-p(F_{tot})}
\right\} \ .
\label{40}
\end{equation}
This expression remains a decent interpolation even if 
$v_y\tau$ is not of the form $(3/2+N)L$.
Symmetrically, for $F_{tot}<0$ the average 
traveling distance is $-\Delta y (\tau, -F_{tot})$,
implying for the net average current (\ref{20}) the approximation
\begin{equation}
\langle\dot y\rangle = 
\frac{\int_0^\infty d\tau \, \rho(\tau) [\Delta y (\tau, A+F)- \Delta y (\tau,A-F)]}
{2\, \int_0^\infty d\tau \, \rho(\tau)\, \tau}\ ,
\label{50}
\end{equation}
where $\pm A$ are the two states of the dichotomous noise 
$\xi_y(t)$ and $\rho(\tau) = \gamma e^{-\gamma\tau}$ is the
distribution of sojourn times, i.e. $\gamma$ is
the flip rate.

The agreement of our analytic prediction (\ref{50}) with 
the simulations in Fig. 1b is remarkably good
in view of the various underlying approximations.
In particular, our assumption that 
the particle covers at least a distance $3L/2$
during the time $\tau$ renders (\ref{40}) doubtful
unless $v_y\tau=F_{tot}\tau/\eta>3L/2$.
To fulfill this condition for all the forces 
$F_{tot}=A\pm F$ notably contributing in (\ref{50})
requires that $A-|F| > 3\gamma\eta L/2$.
Indeed, for $A < 3\gamma\eta L/2$
ANM is found to disappear in numerical 
simulations of (\ref{10}).

On the other hand, ANM is expected to subsist for numerous 
generalizations of our original model (\ref{10}).
Fig. 2 exemplifies a
setup with a 2-dim. array of obstacles.
For symmetry reasons, the current (\ref{20}) remains exactly
the same as in Fig. 1, but the parallelization now admits 
to simultaneously transport a much larger number of particles.
Such a device can be readily realized by a modification of
those studied experimentally in \cite{exp} and theoretically
in \cite{the}. We emphasize that while these modifications
of the experimental setups are straightforward, the physics,
however, is completely different.
A further step towards a realistic experimental system
is achieved by choosing
\begin{equation}
\xi_y(t)=\xi_{th}(t)+f(t) \ ,
\label{60}
\end{equation}
where $\xi_{th}(t)$ is another thermal white noise 
like $\xi_x(t)$ (but statistically independent) 
and $f(t)$ switches {\em periodically} between $\pm A$
with $\rho(\tau)=\delta(\tau-\tau_{ac})$.
For not too weak driving $f(t)$ and with the appropriately
adjusted definition $F_{tot}:=f(t)+F$, the corrections in
(\ref{30}), (\ref{40}) due to the thermal noise $\xi_{th}(t)$ are
small and thus (\ref{50}) remains a valid approximation
provided $A-|F| > 3\eta L/2\tau_{ac}$.

An experimental realization of the
system (\ref{10}), (\ref{60}) along the
lines of \cite{exp,bec}
is presently under construction in the 
labs of C. Bechinger and P. Leiderer.
Henceforth, we focus on such experimentally realistic
parameter values in our quantitative examples, see Fig. 3.
In particular, the agreement of (\ref{50}) 
with the numerical simulations in Fig. 3a
for $\tau_{ac}=1\, s$
is again rather good in the parameter range 
$|F| < 0.14\, pN$ compatible with 
$A-|F| > 3\eta L/2\tau_{ac}$.

In (\ref{40}), we have completely neglected the
possibility that a trapped particle may escape 
from the trap due to the ambient thermal noise.
This is justified as long as $\tau$ 
is much smaller than the mean escape
time $\tau_{esc}(F_{tot})$ out of a trap. 
Turning to the opposite case $\tau\gg\tau_{esc}(F_{tot})$,
we start by calculating the time which a particle
needs to advance by one period $L$ along the $y$-axis for $F_{tot}>0$:
This time is approximately $L/v_y$ if the trap within such a 
period is avoided and $L/v_y+\tau_{esc} (F_{tot})$ otherwise.
The respective probabilities are $p$ and $1-p$, 
approximated by (\ref{30}), i.e. the average time to cover
one period $L$ is $p\, L/v_y + [1-p]\, [L/v_y+\tau_{esc} (F_{tot})]$.
The resulting average traveling distance during the 
(large) time $\tau\gg\tau_{esc}(F_{tot})$ is
\begin{equation}
\Delta y(\tau, F_{tot})=
\frac{\tau \,L}{\frac{L\eta}{F_{tot}}+[1-p(F_{tot})]\tau_{esc}(F_{tot})} \ .
\label{70}
\end{equation}
Thus, if those large $\tau$ dominate, $\Delta y/\tau$ becomes independent
of $\tau$ and (\ref{50}) can be rewritten as
\begin{equation}
\langle\dot y\rangle = \frac{1}{2}\left[
\frac{\Delta y (\tau, A+F)}{\tau}- \frac{\Delta y (\tau,A-F)}{\tau}\right] \ ,
\label{80}
\end{equation}
independent of $\rho(\tau)$.
For small $F_{tot}>0$, the first term in the denominator of (\ref{70})
dominates and hence $\Delta y$ {\em increases} in the expected linear
response manner with increasing $F_{tot}$.
As $F_{tot}$ gets larger, $1-p$ approaches $1$ (cf. (\ref{30}))
and the escape time 
$\tau_{esc}$ increases very fast (cf. (\ref{90}) below), 
implying the existence of a maximum
and a subsequent {\em decay} of $\Delta y$.
As a consequence of this increasing ``stickiness'' of the traps
with increasing $F_{tot}$ \cite{ban72,bal95},
we recover once again ANM in (\ref{80}) 
provided $A$ is sufficiently large.

Focusing on the model (\ref{10}), (\ref{60}),
one can approximate $\tau_{esc}(F_{tot})$ by the mean first passage 
time from $x=0$ to $x=b/\sin\theta$ of the auxiliary 1-dim. dynamics 
$\eta \dot x(t)=- F_{tot}\cos\theta + \xi_x(t)$ 
with a reflecting boundary at $x=0$,
reading \cite{han90}
\begin{eqnarray}
& & \tau_{esc}(F_{tot}) 
= \frac{b^2\, \eta}{k_BT}\,\frac{e^\alpha -\alpha-1}{\alpha^2 \sin^2\theta}
\nonumber\\
& & \alpha := b\, F_{tot}\cot\theta/k_BT \ .
\label{90}
\end{eqnarray}
The agreement of (\ref{70})-(\ref{90})
with the numerical simulations in Fig. 3a for $\tau_{ac}=25\, s$
is quite satisfactory. In particular, the predicted ANM 
is indeed recovered.

We emphasize that the basic 
physical origins of ANM are completely different
in the small-  and large-$\tau$ regimes
as quantified by (\ref{40}), (\ref{50}) and (\ref{70}), (\ref{80}),
respectively:
In the former case, 
escapes out of the traps are negligible, while
transient ``first-trapping events'' after each jump of 
$F_{tot}$ provide the crucial mechanism for ANM.
In the latter case, these transients are negligible, while the 
``re-escape events'' are now at the origin of ANM.
This remarkable feature that two completely different physical
mechanisms both support one and the same phenomenon,
suggests that ANM will also be present in the so far
disregarded intermediate-$\tau$ regime.
Furthermore, on the basis of our above physical insight an
immediate educated guess is to add up (\ref{40}) and
(\ref{70}) and then evaluate (\ref{50}). Both these
predictions are nicely confirmed by Fig. 3b.

A more sophisticated analysis can again be
based on our usual assumption that a particle 
always closely passes by the leftmost edge of any obstacle 
attached to the ``right corridor wall'' when $F_{tot}>0$,
as indicated in Fig. 1a.
Consequently, the traveling times through any period $L$  
are governed by one and the same probability
distribution $\psi (t)$, independent of the 
particle's past (Markov property).
Similarly as in (\ref{70}), this distribution
is approximately given by
\begin{equation}
\psi(t)=p\,\delta(t-\tau_1)+ 
(1-p)\,\Theta (t-\tau_1) 
\frac{e^{-(t-\tau_1)/\tau_{esc}}}{\tau_{esc}} \ ,
\label{110}
\end{equation}
where $\Theta(t)$ is the Heaviside function and $\tau_1:= L/v_y$. 
In this way, the original
2-dim. problem can be approximately reduced
to a 1-dim., uni-directional hopping process 
characterized by $\psi(t)$.
Such processes have been analyzed in detail in the context
of renewal theory \cite{cox}.
Along these lines, we obtain for the Laplace transformed displacement
$\Delta \tilde y(s,F_{tot})
:=\int_0^\infty dt\, \Delta y(t,F_{tot})\, e^{-ts}$
the result
\begin{equation}
\Delta \tilde y(s,F_{tot})=\frac{L}{s}\,
\frac{\tilde\psi(s)}{1-\tilde\psi(s)}\,
\frac{e^{\tau_1 s} -1}{\tau_1 s}
\label{100}
\end{equation}
where $\tilde \psi(s)$ is the Laplace transform of $\psi(t)$.
While the first two factors on the right hand side of (\ref{100})
are well known \cite{cox}, the last factor accounts for the fact 
that the particle actually proceeds
continuously rather than in discrete jumps of length $L$.
Moreover, after the Laplace back-transformation of (\ref{100}), 
a final transformation
$\Delta y(\tau,F_{tot})\mapsto \Delta y(\tau -3L/2v_y,F_{tot}) + 3L/2$
is required since the ``basic distance'' $3L/2$,
which the particle covers before encountering the first trap
(see Fig. 1a and below (\ref{30})),
is not yet taken into account by (\ref{100}).
For very small and large $\tau$-values
one then recovers 
our previous results (\ref{40}) and
(\ref{70}), respectively, while
for more general $\tau$-values, a numerical evaluation 
of the Laplace back-transformation is necessary.
A more detailed derivation of these analytical results
will be presented elsewhere.
A typical example is depicted in Fig. 3b, in good agreement 
with the numerical simulations.

In conclusion, we have demonstrated that a single, classical
Brownian particle in a periodic, symmetric 2-dim. potential 
landscape can exhibit the {\em prima facie} quite astonishing 
phenomenon of absolute negative mobility under suitable far 
from equilibrium conditions.
In general, the effect is simultaneously supported by two 
completely different physical mechanisms and, in contrast to \cite{ban72},
is not restricted to adiabatically slow non-equilibrium perturbations.
The phenomenon is moreover robust against modifications of the potential.
It can occur in practically any potential landscape that provides
``traps'' with increasing ``stickiness'' as external force strengths increase.
The setup discussed here is particularly suitable for an experimental
realization along the lines of \cite{exp,the,bec}.

This work has been supported by the
DFG-Sachbeihilfe HA1517/13-4 and the 
Graduiertenkolleg GRK283.

\vspace*{-0.2cm}

\begin{figure}[h]
\centerline{\epsfxsize=8.5cm
\epsfbox{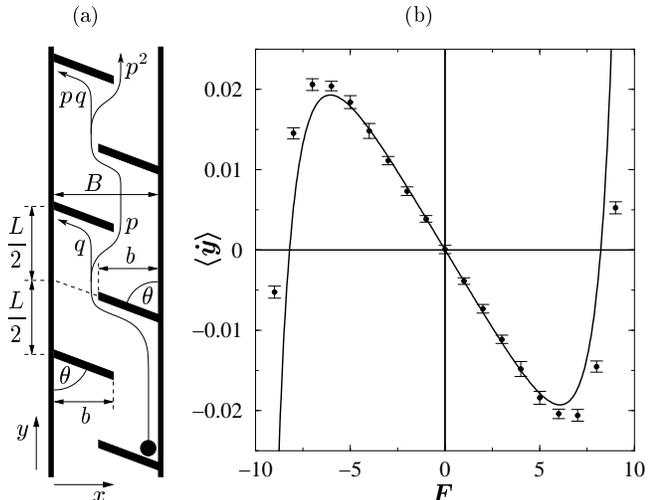}}
\caption{(a): 
Hard-wall potential $V(x,y)$, defined as zero in the white 
and infinity in the black regions. It confines the motion
of a particle to a ``corridor'' of width $B$ along the 
$y$-axis (the white regions outside this corridor are of
no interest). The black obstacles are arranged in a symmetric,
periodic manner ($180^\circ$ rotation symmetry)
and are characterized by the parameters $\theta$ and $b$
(their extension in $y$-direction is neglected).
Note that $b>B/2$.
Several typical traveling routes 
of a particle under the action of a static force $F_{tot}>0$ 
along the $y$-axis 
are indicated as arrows 
together with their probabilities ($q:=1-p$).
(b):
The current (\protect\ref{20}) versus the static force $F$ in (\protect\ref{10})
for the potential $V(x,y)$ from (a),
Gaussian white noise $\xi_x(t)$, and a dichotomous noise $\xi_y(t)$
that flips at a rate $\gamma$ between $\pm A$.
Using dimensionless units, the parameters values are
$\eta = 1$,
$L=1$,
$B=1$,
$b=0.55$,
$\theta=45^o$,
$k_BT=0.1$,
$A=10$,
$\gamma=0.4$.
Dots with error bars: numerical simulations of (\protect\ref{10}).
Solid line: Analytic approximation (\protect\ref{30})-(\protect\ref{50}).}
\label{fig1}
\end{figure}
\begin{figure}[h]
\centerline{\epsfxsize=8.5cm
\epsfbox{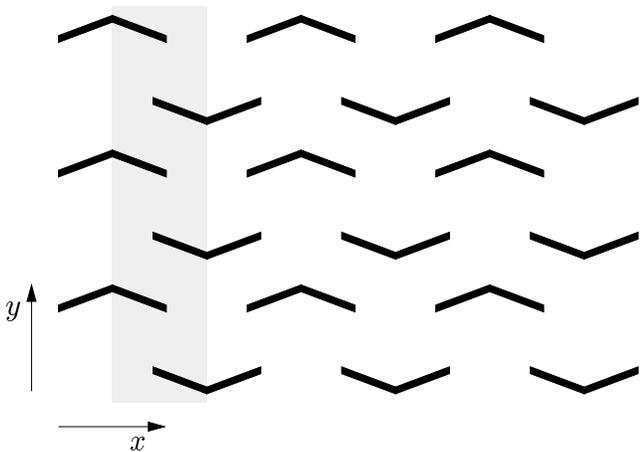}}
\caption{Hard-wall potential $V(x,y)$ like in Fig. 1a
but now periodically continued along the $x$-axis, resulting
in a 2-dim. array of obstacles (``sieve'').
For symmetry reasons, the $y$-component of the
the Brownian motion (\protect\ref{10}) is completely
independent of whether the grey shaded
``corridor'' is endowed 
with perfectly reflecting ``side-walls'' (Fig. 1a) 
or not.}
\label{fig2}
\end{figure}
\begin{figure}[h]
\centerline{\epsfxsize=8.5cm
\epsfbox{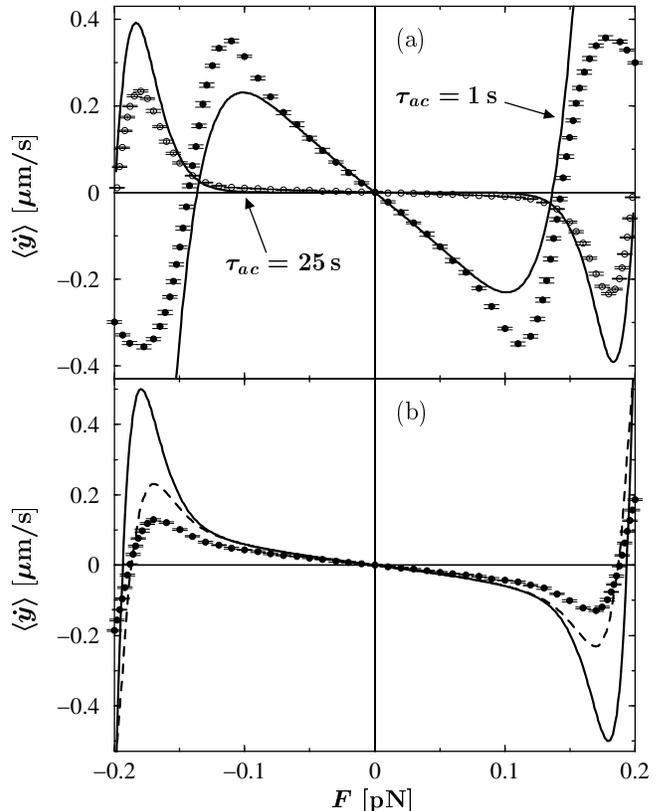}}
\caption{The current (\protect\ref{20}) versus the static force $F$ in (\protect\ref{10}), (\protect\ref{60})
for the potential $V(x,y)$ from Fig. 1a or Fig. 2,
independent Gaussian white noises $\xi_x(t)$ and $\xi_{th}(t)$, 
and a periodic driving $f(t)$ that jumps between $\pm A$
with time constant $\tau_{ac}$.
The parameters values are
$L=4\, \mu m$,
$B=3\, \mu m$,
$b=1.2\, \mu m$,
$\theta=70^o$,
$T=293^o\, K$,
$A=0.2\, pN $.
The Brownian particle is assumed as spherical with radius $r=0.5\, \mu m$
and as subjected to Stokes friction $\eta = 6\pi\nu r$, where $\nu$ is the
viscosity of water.
Dots with error bars: numerical simulations of (\protect\ref{10}), (\protect\ref{60})
with $\tau_{ac}=1\, s$ and $\tau_{ac}=25\, s$ in (a) 
and with $\tau_{ac}=5\, s$ in (b).
Solid line in (a) with $\tau_{ac}=1\, s$: 
Analytic approximation (\protect\ref{30})-(\protect\ref{50}) 
with $\rho(\tau)=\delta(\tau-\tau_{ac})$.
Solid line in (a) with $\tau_{ac}=25\, s$: 
Analytic approximation (\protect\ref{30}), (\protect\ref{70})-(\protect\ref{90}) 
with $\tau=\tau_{ac}$.
Solid line in (b): 
Analytic approximation by adding up (\protect\ref{40}) and (\protect\ref{70})
and then evaluating (\protect\ref{50}) with (\protect\ref{30}), (\protect\ref{90}), 
$\rho(\tau)=\delta(\tau-\tau_{ac})$, and $\tau_{ac}=5\, s$.
Dashed line in (b): 
Analytic approximation for $\tau_{ac}=5\, s$
based on (\protect\ref{110}), (\protect\ref{100}),
as described in more detail in the main text.
Note that all analytical result are based on the assumption of
a point particle. The finite particle radius $r$ has been 
approximately accounted for by replacing $B$ by $B-2r$ 
in (\protect\ref{30}).}
\label{fig3}
\end{figure}


\begin{references}
\bibitem{f1}
After drifting for a time $t$ along the $y$-axis with speed 
$v_y=F_{tot}/\eta$, the 
thermal diffusion along the $x$-axis is approximately 
captured (for not too large $t$)
by a Gaussian distribution with variance $\sigma^2=2Dt$.
With Einstein's relation $D=k_BT/\eta$ and observing
that neighboring obstacles have an overlap $2b-B$
(in $x$-direction) and an approximate 
distance $L/2$ (in $y$-direction) eq. (\ref{30}) follows.
\end{references}
\end{document}